\documentclass[twocolumn,trackchanges]{aastex7}

\usepackage{gensymb}
\usepackage{amsmath}

\newcommand{\Hinout}{\tilde{\mathcal{H}}_\mathrm{in}^\mathrm{out}}
\newcommand{\Hnstar}{\mathcal{H}_N^\star}
\newcommand{\Hnstartilde}{\tilde{\mathcal{H}}_N^\star}
\newcommand{\Hndisktilde}{\tilde{\mathcal{H}}_N^\mathrm{disk}}
\newcommand{\Hntilde}{\tilde{\mathcal{H}}_N}

\begin{document}

\title{Secular Excitation of Polar Neptune Orbits in Pure Disk-Planet Systems}

\author[0000-0002-9305-5101]{Luke B. Handley}
\altaffiliation{NSF Graduate Research Fellow}
\affiliation{Department of Astronomy, California Institute of Technology, Pasadena, CA 91125, USA}
\email[show]{lhandley@caltech.edu}  

\author[0000-0002-7094-7908]{Konstantin Batygin}
\affiliation{Division of Geological and Planetary Sciences, California Institute of Technology, Pasadena, CA 91125, USA}
\email{kbatygin@caltech.edu}

\begin{abstract}

The stellar spin-orbit angles of Neptune-sized planets present a primordial yet puzzling view of the planetary formation epoch. The striking dichotomy of aligned and perpendicular orbital configurations are suggestive of obliquity excitation through secular resonance---a process where the precession of a hot Neptune becomes locked onto a forcing frequency, and is slowly guided into a perpendicular state. Previous models of resonant capture have involved the presence of companion perturbers to the star-planet-disk system, but in most cases, such companions are not confirmed to be present. In this work, we present a mechanism for exciting Neptunes to polar orbits in systems without giant perturbers, where photo-evaporation is the self-contained mechanism. Photo-evaporation opens a gap in the protoplanetary disk at $\sim$1~au, and the inner disk continues to viscously accrete onto the host star, precessing quickly due to the perturbation of the outer disk. As the inner disk shrinks, it precesses more slowly, and encounters a resonance with the $J_2$ precession of the Neptune, quickly exciting it to a polar configuration. While likely not applicable to more massive planets which trigger back-reactions onto the disk, this mechanism reproduces the obliquities of small planets in multiple respects.

\end{abstract}


\section{Introduction} \label{sec:Introduction}

For over two centuries—since the earliest formulations of the Nebular Hypothesis \citep{Kant1755,Laplace1796}—the expectation of coplanarity has anchored our understanding of how planetary systems form. Modern models inherit this assumption: planets emerge from a thin, dissipative disk whose geometry imprints a common orbital plane. Deviations from co-planarity signal additional processes beyond conglomeration itself. In this framework, the spin–orbit angle between an exoplanet’s orbit and its host star’s rotation axis (often called the stellar obliquity, $\psi$) serves as a vital probe of mechanisms that operate concurrently with planet formation. 



Measurements of exoplanet spin-orbit angles quickly followed discoveries of the first transiting planets. Indeed, the first exoplanet system to be successfully observed in transit, HD~209458 \citep{Henry2000,Mazeh2000,Charbonneau2000}, also permitted the first detection and modeling of spectral line-profile distortions for a planet \citep{Queloz2000,Bundy2000,Snellen2004,Winn2005}, the so-called Rossiter McLaughlin (RM) effect first discovered during transits in binary star systems \citep{Rossiter1924,McLaughlin1924}. The spin-orbit alignment of HD~209458~b along with contemporary transiting systems (e.g., HD~149026~b \citep{Wolf2007} and HD~189733~b \citep{Winn2006} were the next two detections) cemented the assumption of co-planarity as standard.

With time, though, the growing obliquity census unearthed puzzling observations. XO-3~b \citep{Hebrard2008,Winn2009}, HD~80~606~b \citep{Moutou2009}, and WASP-14~b \citep{Johnson2009} all demonstrated orbits misaligned from their hosts by $\gtrsim$~40$\degree$. The planet HAT-P-7~b was even found to reside in a retrograde, near-polar ($\psi\sim$90$\degree$) orientation \citep{Winn2009}. These early detections of misalignment led to the immediate interpretation that they stemmed from a process not canonical to planet formation itself---namely post-nebular migration mechanisms active for a subset of planetary systems \citep{Fabrycky2009}. Examples of such processes include the von~Zeipel-Kozai-Lidov (ZKL) induced tidal migration mechanism \citep{Fabrycky2007,Wu2007}, planetary scattering events \citep{Chatterjee2008}, combinations of the two \citep{Nagasawa2008}, and secular chaos \citep{Wu2011}. Consequently, discussions of stellar obliquities shifted toward weighing the relative frequencies of these exotic migration channels versus those driven by disk interactions \citep{Morton2011}, rather than reconsidering the foundational assumption of coplanarity that underpinned the field’s expectations.

As the number of measurements grew large, it became clear that our interpretation of these independent modes of evolution is even further obscured by \textit{stellar} physics. \cite{Triaud2010} presented an even split of aligned and misaligned planets in their seminal sample of hot Jupiters, and ultimately found no distinguishing features among the misaligned planets. \cite{Winn2010} was the first to recognize that the differentiating characteristic was in fact a stellar feature, the stellar effective temperature, which indicates the nominal location of the radiative zone in the stellar interior. Their analysis showed that cool stars tend to be aligned with the orbits of hot Jupiters, while hot stars are often misaligned. The so-called ``$\lambda - T_\mathrm{eff}$" relationship (where $\lambda$ indicates the sky-projection of the stellar obliquity angle) remains the most tried and true observable for stellar obliquities, with $\sim$200 measurements as validation today (see, e.g., the TEPCat database\footnote{https://www.astro.keele.ac.uk/jkt/tepcat/obliquity.html} from \cite{Southworth2011}). 

Commonly, the ``$\lambda - T_\mathrm{eff}$" relation is discussed in the context of stellar tidal mechanisms which might act to diminish spin-orbit measurements over time (e.g., equilibrium tides \citep{Zahn1997,Zahn2008}, inertial tides \citep{Ogilvie2009}, and resonant locking \citep{Zanazzi2024}). The timescales of such effects are strongly dependent on the mass of the planet. The alignment of hot Jupiters around cool stars suggests they are massive enough to realign their hosts. However, hot Neptune planets have an order of magnitude less angular momentum, and thus, these realignment effects are likely not significant. 




This idea is corroborated by observations, as the distribution of Neptune spin-orbit angles around cool stars is distinct from that of giant planets. They appear to demonstrate dichotomy of orientations, where one population appears consistent with alignment, but the other is dramatically inclined into polar orientations \citep{Knudstrup2024,EspinozaRetamal2024,Handley2025}. Intermediate spin-orbit misalignments are not observed. Dynamical explanations are presented on a case-by-case basis, but fail to explain the ensemble. More specifically;

\begin{itemize}
    \item The ZKL channel predicts a multi-modal distribution of misalignments (without preference for 90$\degree$) and requires either a mutually inclined \citep{Fabrycky2007} or eccentric \citep{Naoz2013} massive companions. Chaotic evolution of the stellar spin axis during migration can further complicate the picture \citep{Storch2014}.
    \item Nodal precession \citep{Yee2018,Xuan2020} produces a smooth distribution of obliquities, even under the most favorable orientations of outer perturbers \citep{Rubenzahl2021}.
    \item Planet-planet scattering events \citep{Chatterjee2008} tend to increase the inclinations of the inner scattered body, but rarely to larger than $\sim$60$\degree$. Secular chaos \citep{Wu2011} faces a similar problem, as it does not preferentially excite highly inclined orbits.
    \item Protoplanetary disk torquing \citep{Batygin2012,Spalding2014} can generate polar orbits, but predicts a relatively broad obliquity distribution and requires a primordial stellar companion.
\end{itemize}

Rather than being a consequence of the aforementioned mechanisms, we argue that the $0\degree-90\degree$ dichotomy is a smoking gun for secular resonance encounters. Secular resonance can ensue when the planet's orbit is perturbed by an inclined, slowly time-varying external potential. If the effective nodal precession frequency of the perturbing potential starts greater than that of the planet's orbit and slows with time, capture into a resonance can occur, locking the system into a state where the lines of node become co-linear. Further slow-down of the inherent precession frequency of the planet results in an amplified obliquity. 

Two applications of this resonant mechanism have already been explored. \cite{Batygin2016} showed that small super-Earth planets with interior hot Jupiters can be boosted to polar orbits due to the diminishing precession induced by stellar oblateness. \cite{Petrovich2020} invoked an external giant perturber and a decaying transition disk to excite hot Neptunes to polar orbits. We note that these mechanisms (along with our own) can be reduced to a mathematically indistinguishable model. However, the origin of the perturbing frequency is significant. Both of the above applications necessitate an internal or external massive planet to impose the resonance. Among the hot Neptunes with polar orbits, there is little evidence of the occurrence of such perturbers. WASP-107 \citep{Piaulet2021} and HAT-P-11 \citep{Yee2018}, while foundational examples to the field, are the only systems which clearly fit this dynamical mold.

To summarize, the surge of observational studies on planet–star (mis)alignments over recent decades has challenged the once-trivial assumption of coplanarity. Small planets, which contribute negligibly to the overall gravitational potential, serve as tracers of a system’s dynamical evolution during its formative epoch. However, the stark disparity between observed architectures and theoretical expectations suggests that our understanding may be incomplete. The lack of sustained perturbing bodies upholds this idea, and suggests a primordial origin for obliquity excitation.

In this paper, we present a distinct mechanism which explains stellar obliquity observations assuming only an \textit{isolated} hot Neptune planet in a pure disk-star-planet system. In our analysis, photo-evaporation is the self contained mechanism which permits the opening of a gap in the early protoplanetary disk at $\sim$1~au. The outer disk acts as a large perturbing body that induces rapid nodal precession on the inner (which need only be very slightly mutually inclined). The inner viscously accretes while the outer is evaporated over a longer timescale. Critically, the inner shrinks towards the star, which causes the precession rate to slow until eventual commensurability with the planet's precession rate. This model reproduces the aforementioned resonant mechanism and excites hot Neptunes into $90\degree$ orbits for reasonable choices of disk parameters.

Our paper is organized as follows. In Section \ref{sec:Model} we introduce the model and discuss our assumptions. In Section \ref{sec:DiskDisk} we compute the evolution of the inner and outer disk system, and then compute the resulting behavior of the hot Neptune in Section \ref{sec:PlanetDisk}. In Section \ref{sec:Discussion}, we discuss the scope of our model and compare it to observations, and in Section \ref{sec:Conclusion} we conclude.

\section{Dynamical Model} \label{sec:Model}

We begin by considering the evolution of a decaying, self-gravitating protoplanetary disk around an isolated star. During the initial few million years, disk mass loss is primarily governed by accretion, whether by turbulent angular momentum transport (for example, the magnetorotational instability (MRI; \cite{Balbus1991}), vertical shear instability (VSI; \cite{Nelson2013}), etc.), or by magnetohydrodynamic (MHD) disk winds \citep{Lesur2021}. However, as the disk drains over time, mass loss due to photo-evaporative processes becomes increasingly significant \citep{Alexander2006}. At sufficiently large radii, ionizing radiation\footnote{Hydrodynamic simulations in \cite{Owen2011} indicate the relevant energies are predominantly X-rays, rather than the UV radiation supposed in \cite{Alexander2006}} from the central star heats particles on the surface of the disk, granting sufficient thermal energy to escape the local gravitational potential as a photo-evaporative wind \citep{Clarke2001}. This transition in the dominant mass-loss mechanism leads to the opening of a gap in the disk, typically between 1 and 5~au. In this phase, the inner disk continues to accrete onto the star under viscous forces, while the outer disk is gradually eroded by photo-evaporation \citep{Liu2022}. While the end state of this mechanism is a class of so-called `transition disks' which are well corroborated by observations \citep{Espaillat2014}, we will show that the dynamical evolution during this short gap-opened phase can have significant consequences.

To accomplish this, we will make use of the angular momentum hierarchy among our system constituents to simplify the problem, and consider only the dominant dynamics at each stage. We assume a continuous surface density profile like that in \cite{Bell1997} and \cite{Andrews2009}:

\begin{equation} \label{eq:sigma}
    \Sigma = \Sigma_0 \left( \frac{a_0}{a} \right)^{1/2},
\end{equation}
where $a_0= 1$~au is a reference size scale for the disk, and $\Sigma_0=1500$ g cm$^{-2}$ is the surface density at that scale radius. While our choice of power law reflects that of a constant accretion rate and optically thin disk \citep{Armitage2010,Lambrechts2017}, we chose such a power law for analytic simplicity, and it is not a necessary condition. The angular momentum of a differential disk annulus is then 
\begin{equation} \label{eq:angmom}
    d\mathcal{L}=2\pi\Sigma\sqrt{GMa^3}da, 
\end{equation}
where $G$ is the gravitational constant and $M$ is the mass of the host star (assumed 1 M$_\odot$), and we may compute the angular momentum of the inner and outer disks by integrating over the semi-major axis spanned by each. The inner disk will be bounded between the magnetic truncation radius $a_X$ which we take to be $0.1$~au \citep{Bouvier2007} and the radius at which the gap opens $A_\textrm{gap}$ at $\sim$1~au. 


The outer disk then ranges from $A_\mathrm{gap}$ to infinity (an outer radius of $\sim$100~au is observationally consistent \citep{Andrews2009}, but mathematically indistinguishable in this limit). By integrating over each disk annulus, we find $\mathcal{L}_\textrm{out}/\mathcal{L}_\textrm{in}>10^3$, implying that the outer disk completely dominates the dynamics of the inner, and that we may neglect any back-reaction between the two. 

The same relationship should also be validated for the inner disk-planet system. The orbital angular momentum of a Neptune-sized planet on a circular orbit is given by $\mathcal{L}_N = M_N\sqrt{GMa_N}$, where $a_N$ is the orbital radius of the planet. Our regime of interest is $a_N\sim$0.05~au, and we find that $\mathcal{L}_\textrm{in}/\mathcal{L}_\textrm{N}\sim$40, ensuring that the Neptune is analogously dominated by the perturbation of the inner disk, and that we may neglect the back-reaction from from the planet. We further remark that $\mathcal{L}_\textrm{in}/\mathcal{L}_\textrm{N}\gtrsim10$ for an inner disk extending down to $\sim$0.35~au, implying that our assumption is robust even as the inner disk shrinks throughout the early accretionary stage\footnote{We note that the angular momentum hierarchy may not be satisfied if the inner disk surface density were depleted compared to our model, which is discussed in Section \ref{sec:caveats}.}. Therefore, we may decouple the problem into an investigation of the mutual disk system, and allow the behavior of the inner disk to then solely dictate the evolution of the planet's orbit.

\subsection{The Gap Opened Disk} \label{sec:DiskDisk}

We begin by examining how the outer disk drives the inner disk’s dynamical evolution. We choose coordinates aligned with the outer disk, so the inner has a small inclination $i_\textrm{in}$ and the outer lies at $i_\textrm{out}=0$. Because the two components can have nearly identical semi-major axes (coincident at $A_\textrm{gap}$ before viscous evolution) and differ by only a small mutual inclination\footnote{The minor inclination can originate from a range of effects, including the cluster tidal potential, a primordial stellar binary, stellar flyby, etc.} of $\sim$1$\degree$, we adopt a Laplace-Lagrange formalism \citep{Murray1999}. Laplace-Lagrange permits analytic solutions to the secular behavior of the inner disk through use of an approximate integrable Hamiltonian where the strengths of interactions are parametrized by the Laplace coefficients $b^{(m)}_{s}$, which are functions of $\beta= a_\textrm{out}/a_\textrm{in}$, the ratio of the semi-major axes of an outer and inner body, respectively. However, the standard Laplace coefficients are not well suited for the problem at hand, due to a singularity at $\beta=1$ (two overlapping orbits will have an infinitely strong secular perturbation on one another). To remedy this shortcoming, we adopt the softened Laplace coefficients of \cite{Hahn2003} which account for the finite thickness of each disk component to mitigate numerical divergence during the early dispersal stage. For any disk annulus $a_\textrm{in}$ in the inner disk, we may compute the Laplace coefficient of the perturbation due to any outer disk annulus $a_\textrm{out}$ using

\begin{equation} \label{eq:laplace}
\begin{split}
    \tilde{b}_s^{(m)}(\beta,\mathfrak{h}_\mathrm{in},\mathfrak{h}_\mathrm{out}) 
    =& {} \\
    \frac{2}{\pi} \int_{0}^{\pi} \frac{\mathrm{cos}(m\phi)\,d\phi}{\left\{ (1+\beta^2)\left[1+\frac{1}{2}(\mathfrak{h}_\mathrm{in}^2 + \mathfrak{h}_\mathrm{out}^2)\right] 
    - 2\beta\,\mathrm{cos}\phi \right\}^{s} }& {},
\end{split}
\end{equation}
where $m$ is the azimuthal mode number, $\phi$ is the true longitude of the perturber, $s$ is a half-integer, and $\mathfrak{h}_\mathrm{in}$ and $\mathfrak{h}_\mathrm{out}$ are the aspect ratios $h/r$ of the inner and outer annuli in question (we assume 0.05 for both). The softened coefficients\footnote{Note that to save space, we will hereafter write $\tilde{b}_s^{(m)}$ rather than $\tilde{b}_s^{(m)}(\beta,\mathfrak{h}_\mathrm{in},\mathfrak{h}_\mathrm{out}) $, with the arguments being implicit.} are valid for values of $\beta$ close to 1, and more correctly encapsulate the three dimensional structure of each annulus.

For any arbitrary inner disk annulus $m_\mathrm{in}$ located at $a_\mathrm{in}$, the differential Hamiltonian due to an outer disk annulus $m_\mathrm{out}$ located at $a_\mathrm{out}$ is \citep{Hahn2003}

\begin{equation}
    d\mathcal{H}^\mathrm{out}_\mathrm{in} = \frac{G dm_\mathrm{in} dm_\mathrm{out}}{4 a_\mathrm{in}}\left(\frac{a_\mathrm{out}}{a_\mathrm{in}}\right)\, \tilde{b}_{3/2}^{(1)}\, \frac{i_\mathrm{in}^2}{2}.
\end{equation}
To derive the equations of motion, we transform this Hamiltonian into the canonical Poincar\'e coordinate system, defined by
\begin{equation} \label{eq:poincare}
    z=-\Omega, \quad Z=1-\mathrm{cos}\,i\approx \frac{i^2}{2},
\end{equation}
which implies a rescaling by the angular momentum of the inner annulus $d\Lambda_\mathrm{in}=dm_\mathrm{in}\sqrt{G M a_\mathrm{in}}$ to remain canonical \citep{Lichtenberg1983}. The Hamiltonian is thus
\begin{equation}
    d\Hinout=\frac{1}{4}\sqrt{\frac{GM}{a_\mathrm{in}^3}}\left(\frac{dm_\mathrm{out}}{M}\right)\left(\frac{a_\mathrm{out}}{a_\mathrm{in}}\right)\,\tilde{b}_{3/2}^{(1)}\, Z_\mathrm{in}.
\end{equation}
Note that $d\Hinout$ is now represented in units of frequency rather than energy. The total Hamiltonian of annulus $m_\mathrm{in}$ is found by integrating over the entire outer disk:
\begin{equation} \label{eq:innerhamiltonian}
\begin{split}
        \Hinout=&\int^\infty_{A_\mathrm{gap}} d\Hinout\\
        &=\frac{\pi}{2}\Sigma_0a_0\sqrt{\frac{G}{M a_\mathrm{in}^5}}\\
        & \quad \times\left[\int^\infty_{A_\mathrm{gap}}\left(a_\mathrm{out}\right)^{3/2}\,\,\tilde{b}_{3/2}^{(1)}\,da_\mathrm{out}\right]Z_\mathrm{in},
\end{split}
\end{equation}
which is independent of $z_\mathrm{in}$, indicating that the inclination of an inner annulus is constant in time. If not for coupling between adjacent disk annuli, each ring $m_\mathrm{in}$ would then regress independently at a rate
\begin{equation} \label{eq:precfreq}
    \frac{dz_\mathrm{in}}{dt} = \frac{\partial\Hinout}{\partial Z_\mathrm{in}}.
\end{equation}

Numerically, we find that for annuli near the gap, the Laplace coefficient dominates Equation~\ref{eq:innerhamiltonian}, yielding $dz_\mathrm{in}/dt\propto a_\mathrm{in}^5$. As a consequence, the regression rate can vary by orders of magnitude across the inner disk.

To capture this behavior in a tractable way, we instead compute an effective precession frequency for the \emph{entire} inner disk. We argue that once a gap is carved the inner disk remains dynamically rigid, as justified by a comparison of relevant timescales. Assuming an $A_\mathrm{gap}$ of 1~au, evaluating Equation \ref{eq:precfreq} at the outermost edge of the inner disk results in a maximum precession frequency of

\begin{equation}
    t_\mathrm{in}= \left(\left.\frac{dz_\mathrm{in}}{dt}\right|_{A_\mathrm{gap}}\right)^{-1} \approx35\, \textrm{yr}.
\end{equation}
Disk coherence is upheld by the propagation of nodal bending waves between disk annuli, which travel at half the sound speed for disks not undergoing rapid accretion \citep{Zanazzi2018}. Thus, the timescale\footnote{This is, in reality, a worst case estimate for the timescale. The propagation speed of bending waves in central regions of the inner disk, where differential warping would be of the greatest concern, are considerably faster than the estimate at $A_\mathrm{gap}$.} of bending wave propagation is given by

\begin{equation}
    t_\mathrm{bw} = \frac{2A_\mathrm{gap}}{c_s} = 2\left(\frac{r}{h}\right)\Omega_\mathrm{orb}^{-1}\approx6\,\mathrm{yr},
\end{equation}
where $\Omega_\mathrm{orb}$ is the orbital frequency of the disk at $A_\mathrm{gap}$. We construct a dimensionless parameter $\epsilon$ which is the ratio of these two timescales
\begin{equation}
    \epsilon = \frac{t_\mathrm{bw}}{t_\mathrm{in}} < 0.2\,,
\end{equation}
and describes the relative importance of differential precession effects (i.e., warping of the disk due to misalignment of nodes). For small $\epsilon$, the disk maintains rigidity due to rapid equilibrating effects dominating differential precession between inner disk annuli \citep{Batygin2018, Zanazzi2018}, and the disk will not warp significantly\footnote{While we validate the rigidity of the disk to simplify our prescription, it is not a necessary condition. Small warps in the outer disk would not have a significant impact on the dynamics.}. Given the timescales of the problem at hand, disk rigidity is a reasonable assumption. Furthermore, the gap grows in size, and the assumption of rigidity is satisfied only better with time.

We model the inner disk as having a shrinking outer extent $A_\mathrm{disk}$ given by
\begin{equation}
    A_\mathrm{disk}(t) = \frac{A_\mathrm{gap}}{\left(1+\frac{t}{t_v}\right)^{3/2}},
\end{equation}
where $t_v$ is the local viscous timescale at $A_\mathrm{gap}$. We chose a value of $\alpha=10^{-4}$ (see Section \ref{sec:caveats}) such that the viscous timescale is $\sim10^6$~yrs. The total precession rate of the inner disk is computed by taking the angular momentum weighted average of the precession rates of each individual annulus (\cite{Epstein-Martin2022}, see also \cite{Larwood1996}). With our surface density prescription, the total precession frequency may be written as
\begin{equation}
    \nu_D(t)\approx\frac{\int_{a_X}^{A_\mathrm{disk}(t)}\frac{dz_\mathrm{in}}{dt}\,a_\mathrm{in}\,da_\mathrm{in}}{\int_{a_X}^{A_\mathrm{disk}(t)}\,a_\mathrm{in}\,da_\mathrm{in}}.
\end{equation}
At times well before the viscous timescale, $A_\text{disk} \gg a_X$ such that $\nu_D$ scales as $A_\text{disk}^{5}$, which is plotted in Figure \ref{fig:contours}.
Thus, as the radial extent of the disk decays, the averaged precession diminishes significantly, enabling the resonant capture mechanism in the next section.


\subsection{Evolution of Short-Period Neptunes} \label{sec:PlanetDisk}

With a prescription for the behavior of the inner disk, we will now consider the dynamics of the Neptune under the influence of the precessing disk and the rotational bulge of the host star. We adopt a model where the host star has mass 1 M$_\odot$, radius 1.3 R$_\odot$, a rotational period of $2\pi/\Omega_\star=$ 5 days, a tidal love number of $k_2\sim0.2$, and a dimensionless moment of inertia $I_0\sim0.21$, which is characteristic of an $n=3/2$ polytrope. To quadrupole order, this is easily accomplished by modeling the star as a point mass surrounded by a wire \citep{Batygin2016} of mass
\begin{equation}
    m_w=\left(\frac{3 M^2 \Omega_\star^2 R^3I_0^4}{4 G k_2}\right)^{1/3}
\end{equation}
and of semimajor axis
\begin{equation}
    a_w = \left(\frac{16 \Omega_\star^2  R^6k_2^2}{9  GMI_0^2}\right)^{1/3}.
\end{equation}
For our choice of parameters, the wire sits at $a_w\approx10^{-3}$~au, significantly smaller than the orbital distance of the Neptune, which simplifies the Hamiltonian prescription notably.

We will continue our use of a Laplace-Lagrange expansion with the caveat that it does not adequately describe the behavior high inclinations. However, it does provide a correct analytic prescription of the resonant capture mechanism during the low inclination phase. In Poincar\'e coordinates, the rescaled Hamiltonian of the Neptune due to the effect of the oblate host star can be simply written as:

\begin{equation}
    \Hnstar = \nu_N^\star  Z_N
\end{equation}
where
\begin{equation}
    \nu_N^\star = \frac{1}{4}\sqrt{\frac{G M}{a_N^3}}\left(\frac{m_w}{M}\right)\left(\frac{a_w}{a_N}\right)\tilde{b}_{3/2}^{(1)}
\end{equation}
is the characteristic precession frequency of the Neptune due to the stellar quadrupole.

Similar to Section \ref{sec:DiskDisk}, we may write the differential Hamiltonian for the Neptune due to the influence of each inner disk annulus $m_\mathrm{in}$. However, in this case, the inner disk inclination $i_\mathrm{in}$ is nonzero, and we include the term which is fourth order in the inclination of the Neptune to capture the non-linear response of the precession frequency at higher inclinations. After again rescaling to Poincar\'e coordinates, it may be written as 

\begin{equation}
\begin{split}
\mathrm{d}\Hndisktilde &= 
\sqrt{\frac{GM}{a_N^3}}\left(\frac{\mathrm{d}m_\mathrm{in}}{M}\right)
\Bigg[
\frac{1}{4}\left(\frac{a_\mathrm{in}}{a_N}\right)\tilde{b}_{3/2}^{(1)}\,Z_N\\
&\qquad -\frac{1}{2}\left(\frac{a_\mathrm{in}}{a_N}\right)\tilde{b}_{3/2}^{(1)}\,
\sqrt{Z_N Z_\mathrm{in}}\,\cos\,\bigl(z_N-z_\mathrm{in}\bigr)\\
&\qquad -\frac{1}{4}\left(\frac{a_\mathrm{in}}{a_N}\right)^2
\left(\frac{3}{8}\tilde{b}_{5/2}^{(2)} + \frac{3}{4}\tilde{b}_{5/2}^{(0)}\right) Z_N^2
\Bigg].
\end{split}
\end{equation}

\begin{figure*} 
    \centering
    \includegraphics[width=\linewidth]{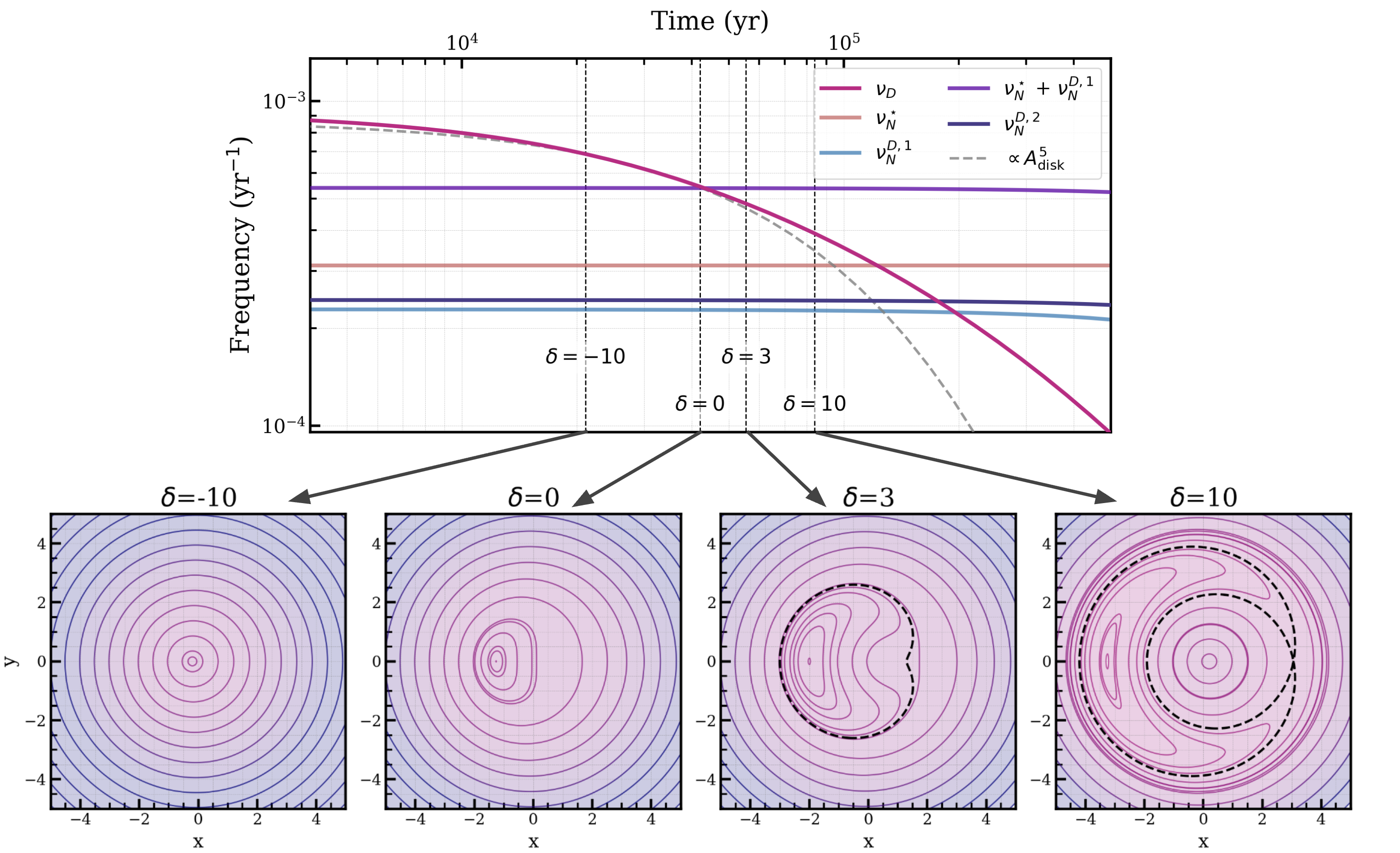}
    \caption{Top: time-evolution of the inner disk and planetary precession frequencies. Bottom: contours of the simplified Hamiltonian (Equation \ref{eq:andoyer}) as functions of the resonance proximity parameter $\delta$, plotted in the canonical cartesian coordinates (x, y) = ($\sqrt{2\Phi}\,\mathrm{cos}\,\phi$, $\sqrt{2\Phi}\,\mathrm{sin}\,\phi$). For $\delta\ll0$ (rapid precession of the inner disk), there is only a single equilibrium point around which all orbits circulate. As $\delta$ increases over time, that equilibrium shifts to higher actions (inclinations) establishing a libration region. When the precession rates of the Neptune and the inner disk are equal ($\delta=0$), the resonance is crossed, and two new equilibria are born at $\delta=3$. The unstable equilibrium lies on the contour which bounds the resonant region, the separatrix, which is plotted in black in the bottom-right two panels. During adiabatic capture, orbits of small action follow the leftward-migrating equilibrium (the crescent shape in the last panel) and remain trapped there for $\delta\gg0$, as $\nu_D$ approaches zero.}
    \label{fig:contours}
\end{figure*}

The total Hamiltonian which describes the evolution of the Neptune's orbit is given by the sum of the contributions from both perturbers, i.e, 
\begin{equation}
\begin{split}
    \Hntilde=
    \Hnstartilde + \int_{a_X}^{A_\mathrm{disk}(t)} d\Hndisktilde \,2\pi a_\mathrm{in}\Sigma\,d a_\mathrm{in}\,.
\end{split}
\end{equation}
The disk-integrated Hamiltonian can be further simplified by defining the following two quantities:

\begin{equation}
    \nu_N^{D,1} = \frac{\pi\Sigma_0}{2}\sqrt{\frac{G}{Ma_N^5}} \int_{a_X}^{A_\mathrm{disk}(t)} a_\mathrm{in}^{3/2}\,\,\tilde{b}_{3/2}^{(1)}\,da_\mathrm{in}
\end{equation}
\begin{equation}
\begin{split}
    \nu_N^{D,2} = &\frac{\pi\Sigma_0}{2}\sqrt{\frac{G}{Ma_N^7}} \\&\quad\times\int_{a_X}^{A_\mathrm{disk}(t)} a_\mathrm{in}^{5/2}\left(\frac{3}{8}\tilde{b}_{5/2}^{(2)} + \frac{3}{4}\tilde{b}_{5/2}^{(0)}\right)da_\mathrm{in}
\end{split}
\end{equation}
Which represent the the precession frequencies induced on the Neptune by the inner disk at second and fourth order in inclination in the expansion, respectively. The values of these frequencies for our model are given in the top panel of Figure \ref{fig:contours}.

\begin{figure}
    \centering
    \includegraphics[width=\linewidth]{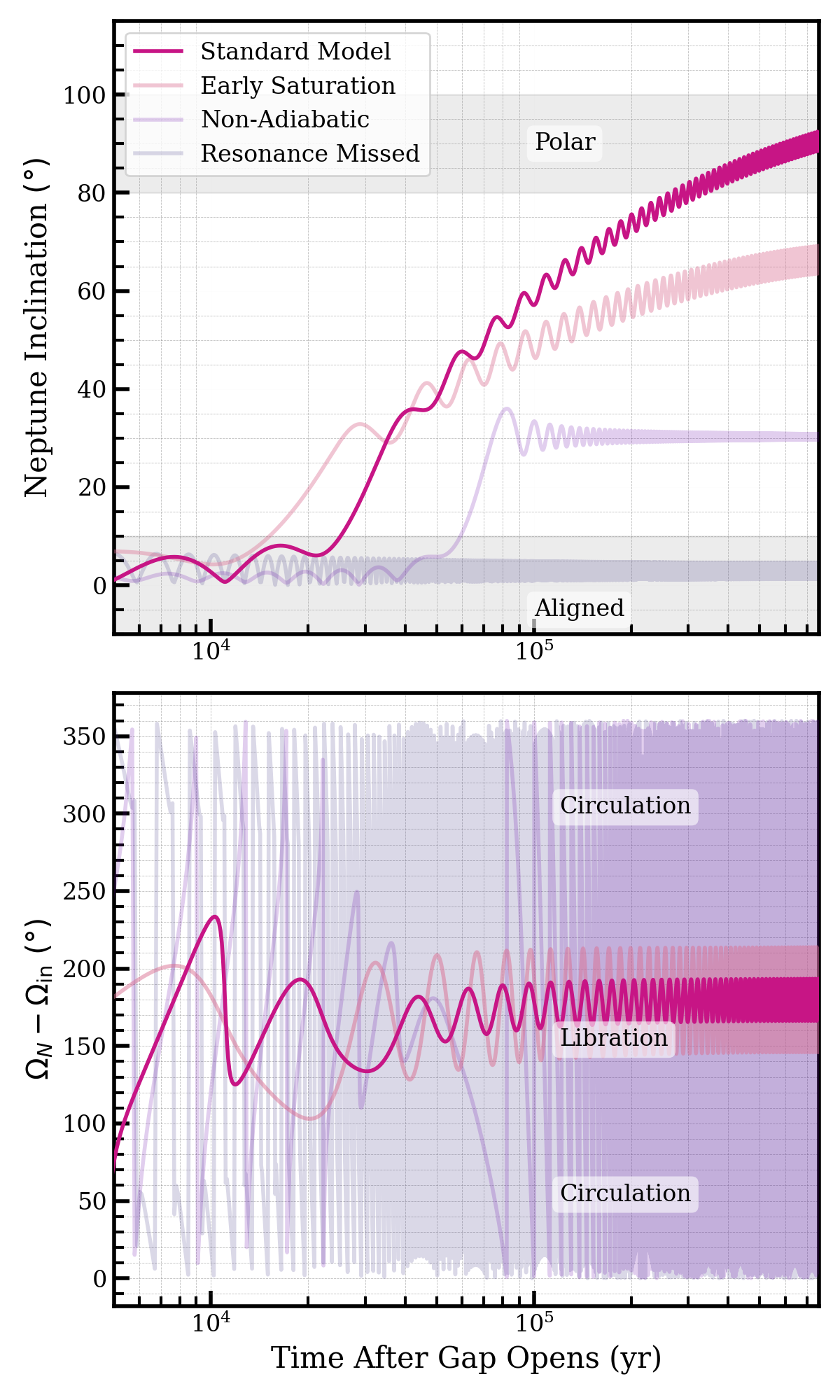}
    \caption{Integrations of the Neptune's Hamiltonian under several informative regimes with\footnote{While integrations below this inclination are often still adiabatic, this value permits many different outcomes by only shifting the truncation radius.} $i_\mathrm{in}=2\degree$ and an initial $i_N=1\degree$. `Standard Model' indicates the expected outcome for realistic inner disk profiles (See Section \ref{sec:Discussion}) as well as our fiducial model with $a_X=0.1$~au. `Early Saturation' indicates an integration with $a_X=0.08$~au that crosses the resonance adiabatically, but a large value of $\nu_N^{D,2}$ shifts the equilibrium to a sub-polar orbit (Equation \ref{eq:saturation}). `Non-Adiabatic' corresponds to $a_X=0.15$~au such that the resonance was crossed, but the adiabatic criterion was not met. `Resonance Missed' was integrated with $a_X=0.06$~au, and results in dynamics dominated by the inner disk such that the resonance is not crossed. Note that the qualitative differences in each case is primarily due to the asymptotic behavior of our power-law prescription of the surface density.}
    \label{fig:integrations}
\end{figure}

Plugging in the precession frequency of the inner disk from the previous section, the total Hamiltonian can now be written as:
\begin{equation} \label{eq:finalham}
    \begin{split}
     \Hntilde &= \bigl(\nu_N^\star+\nu_N^{D,1}\bigr) Z_N - \nu_N^{D,2}\,Z_N^2\\ &- 2\,\nu_N^{D,1} \sqrt{Z_NZ_\mathrm{in}}\,\cos\,\bigl(z_N-\nu_D\,t\bigr).
    \end{split}
\end{equation}
Integrations of Hamilton's equations of motion are found in Figure \ref{fig:integrations}.

To reduce the number of degrees of freedom, we enter a rotating reference frame through use of the type-2 generating function \citep{Morbidelli2002}
\begin{equation}
    \mathcal{G}_2=\bigl(z_N-\nu_D\,t\bigr)\,\tilde{\Phi},
\end{equation}
implying a transformation given by
\begin{equation}
    \begin{split}
        Z_N &= \frac{\partial\mathcal{G}_2}{\partial z_N}=\tilde{\Phi}\\
        \phi&=\frac{\partial\mathcal{G}_2}{\partial \tilde{\Phi}} = z_N-\nu_D\,t\\
        \Hntilde' &= \Hntilde-\frac{\partial\mathcal{G}_2}{\partial t}=\Hntilde-\nu_D\,\tilde{\Phi},\\
    \end{split}
\end{equation}
leading to a Hamiltonian which is now that of a single degree of freedom, and depends on time only through the adiabatic evolution of $\nu_D$. However, it is most practically interpretable by an appropriate rescaling of time and momentum given in \cite{Henrard1983}:
\begin{equation} \label{eq:librationperiod} \tau=\left(\frac{\bigl(\nu_N^{D,1}\bigr)^2\,\nu_N^{D,2}Z_\mathrm{in}}{2}\right)^{1/3}t
\end{equation}
\begin{equation}
        \Phi = \left(\sqrt{2}\,\,\frac{\nu_N^{D,2}}{\nu_N^{D,1}}\right)^{2/3}\tilde{\Phi},
\end{equation}
where $\tau$ describes the libration period of the resonant domain. Finally, we arrive at a prescription of the Neptune's Hamiltonian given by the second fundamental model of resonance \citep{Henrard1983}
\begin{equation} \label{eq:andoyer}
    \Hntilde' = \delta\,\Phi - \Phi^2-2\sqrt{2\Phi}\,\mathrm{cos}(\phi),
\end{equation}
where the parameter $\delta$, which indicates the instantaneous distance to resonance, is given by
\begin{equation} \label{eq:delta}
    \delta=\left(-\nu_D+\nu_N^\star+\nu_N^{D,1}\right)
    \left(\frac{2}{\bigl(\nu_N^{D,1}\bigr)^2\,\nu_N^{D,2}Z_\mathrm{in}}\right)^{1/3}.
\end{equation}
Now, both the conditions for resonant capture and the final inclination state of the Neptune can be understood from only consideration of the time evolution of $\delta$. Contours of this Hamiltonian, which differentiate the regimes of libration (resonance) and circulation, are plotted as functions of $\delta$ in the bottom panels of Figure \ref{fig:contours}.

At early times, the inner disk precesses rapidly ($\nu_D$) due to the small size of the disk gap. If the frequency is large compared to the precession induced on the Neptune by the stellar quadrupole ($\nu_N^\star$) and inner disk ($\nu_N^{D,1}$), then  clearly $\delta\ll0$. After an interval comparable to the viscous timescale passes, the increasingly wide gap between the inner and outer disk tends $\nu_D$ to zero\footnote{Formally, the average precession rate tends to that of an isolated annular ring at the truncation radius $a_X$, dropping several orders of magnitude.} (Equation \ref{eq:precfreq}).  Therefore, at time $t=0$, the condition

\begin{equation} \label{eq:crossing}
    \nu_D>\nu_N^\star+\nu_N^{D,1}
\end{equation}
indicates an inevitable resonance crossing at $\delta=0$, which facilitates the inclination excitation\footnote{While $\nu_N^{D,1}$ can also decrease in time, it does so over a much longer timescale due to the disk depleting from the outside in (Figure \ref{fig:contours}).}. This inequality is critical---we emphasize that the mechanism is entirely insensitive to the initial value of $\nu_D$ so long as it is above this threshold.

Furthermore, given that the Neptune is always assumed to be nearly aligned at $t=0$, the action $\Phi$ is small far from the resonance. Thus, the Neptune may be permanently captured into resonance as long as the evolution is adiabatic, i.e., changes in the topology of the phase-space occur slowly compared to the libration period of the resonant domain \citep{Kruskal1962,Henrard1983,Henrard1993,Friedland2001,Batygin2016}. \cite{Quillen2006} wrote this condition approximately\footnote{The derivative of $\delta$ is a proxy for the ratio of the width crossing time to libration period up to a factor of order unity.} as

\begin{equation} \label{eq:adiabatic}
    \frac{d\delta}{d\tau} = \frac{d\delta}{dt}\left(\frac{d\tau}{dt}\right)^{-1}\lesssim3.
\end{equation}
Note that the definition of $\delta$ used here is slightly different than in \cite{Quillen2006} and some other works (what we call $\delta$ is sometimes left as 3($\delta\pm1$), hence the value of 3 on the right-hand of the inequality). Differentiating Equations \ref{eq:librationperiod} and \ref{eq:delta} gives
\begin{equation} \label{eq:ddeltadtau}
    \frac{d\delta}{d\tau} \approx-\dot{\nu}_D\left(\frac{2}{\bigl(\nu_N^{D,1}\bigr)^2\,\nu_N^{D,2}Z_\mathrm{in}}\right)^{2/3}.
\end{equation}
For a given disk setup (a value of $A_\mathrm{gap}$, a viscosity parameter $\alpha$ \citep{ShakuraSunyaev1973} which gives a viscous timescale, and a surface density profile) these equations establish a minimum value of $Z_\mathrm{in}$ (and thus $i_\mathrm{in}$) for adiabatic evolution. Our numerical integrations of the Hamiltonian give a more relaxed upper bound of 9 for the right-hand side of Equation \ref{eq:ddeltadtau}, resulting in adiabatic capture for a large suite of initial conditions. The forgiving nature of the adiabatic limit is difficult to quantify analytically, but a value greater than 3 is not unexpected (see, e.g., derivations of the adiabatic invariant in \cite{Bellan2008}). For our fiducial parameters, we find that resonant capture is guaranteed for inner disk inclinations as low as $\sim$1$\degree$ for our standard setup.

When capture is achieved, the equilibrium of the Hamiltonian provides an approximate prescription for the end-state action $Z_N$. As $\nu_D$ tends to 0, the Hamiltonian tends to $\Hntilde$ (Equation \ref{eq:finalham}) with $z_N-z_\mathrm{in}=\pi$, and the steady-state equation of motion becomes

\begin{equation} \label{eq:eqofmotion}
    \frac{\partial\Hntilde}{\partial Z_N} = \nu_N^\star +\nu_N^{D,1}\left(1+\sqrt\frac{Z_\mathrm{in}}{Z_N}\right)-2\,\nu_N^{D,2}Z_N=0.
\end{equation}
Following \cite{Batygin2016}, we approximate a solution by expanding as a Taylor series in $\sqrt{Z_\mathrm{in}}$, which to leading order gives
\begin{equation} \label{eq:saturation}
    Z_N \approx \frac{\nu_N^\star+\nu_N^{D,1}}{2\, \nu_N^{D,2}}.
\end{equation}
In our setup, the stellar quadrupolar contribution to the nodal precession dominates the contributions from the inner disk\footnote{In the limit where the stellar quadrupole disappears, polar orientations are still excited if the perturbations from the disk are predominantly from annuli far from the planet (the original derivation in \cite{Batygin2016} explores this limit of the expansion).} giving $Z_N\gtrsim1$ from Equation \ref{eq:saturation}. Any value $Z_N>1$ computed this way is due to the Laplace-Lagrange expansion \footnote{In our Laplace-Lagrange derivation, we have quoted a $\sqrt{Z_N}$ prefactor for the resonant term in Equation \ref{eq:finalham} which holds for small $Z_N$. However, as $Z$ increases in an integration, the prefactor is better approximated as $\sqrt{Z_N(1-Z_N)}$.}, and results in a maximum $Z_N=1$ during real integrations. Thus, the resonant pumping results in a Neptune at a polar configuration for any choice of parameters which gives $\nu_N^\star+\nu_N^{D,1}\geq2\,\nu_N^{D,2}$. Given the short proximity of the Neptune's orbit, we find this easily satisfied. We provide a numerical example for a system which fails to satisfy this requirement and saturates before reaching a polar orbit in Figure \ref{fig:integrations}.

A final point of consideration is whether the Neptune's orbit is stable against eccentricity excitation. At higher orders in the expansion of the Hamiltonian, coupling of inclination and eccentricity (ZKL) could lead to eccentric instabilities that would ultimately detune the resonance if left unchecked. \cite{Petrovich2020} pointed out that the characteristic timescale of these instabilities is likely superseded by apsidal precession induced by General Relativity (GR) for orbits of a few days or less ($P_N\sim4$ days, like we have assumed here), thus `shielding' this class of planets as they evolve with the resonance. 

To validate that this assumption holds for our own setup, we can compare the timescale of destabilization to that of GR-induced precession. As an estimate for the ZKL timescale, we use the distant tides approximation from \cite{Terquem2010}, Equation 9:

\begin{equation} \label{eq:zkl}
    \tau_\text{ZKL} = \frac{(1+n)(1-\eta^{-n+2})}{(-n+2)(-1+\eta^{-n-1})}\frac{A_\text{gap}^3M}{a_N^3M_\text{disk}}\frac{P_N}{2\pi}.
\end{equation}
Here, $n$ indicates the index of the surface density profile ($n$=1/2 in our model), $\eta$ is the ratio
$A_\text{gap}/a_X$, and $M_\text{disk}$ is the total mass of the disk. For comparison, the precession frequency of the argument of periastron caused by GR is given by \cite{Misner1973}

\begin{equation} \label{eq:gr}
    \dot{\omega}_\text{GR}=\frac{3(G M)^{3/2}}{a_N^{5/2}c^2},
\end{equation}
where $c$ is the speed of light. For the Neptune's orbit, this gives a GR timescale of 20,000 years. \cite{Fabrycky2007} showed that this precession suppresses instabilities so long as it not greater than $\sim$2 times the ZKL timescale.

Evaluating Equation \ref{eq:zkl} with our parameters indicates the ZKL timescale is $\sim$7,000 years---just below the necessary condition. Thus, ZKL oscillations would start to destabilize the eccentricity as the planet approaches a polar orbit. However, this susceptibility can again be attributed to the asymptotic behavior of our surface density profile. If we attempt to mitigate this numerical effect by considering $a_X$=0.15~au, the timescale becomes 12,000 years, indicating complete suppression of instabilities. Hence, we argue that a ZKL-active system is ultimately a pathological example, as the structure of a hydrodynamically resolved inner disk would not induce such a strong perturbation. GR is a thus a natural explanation for the presence of polar orbits at short orbital periods.


\section{Discussion} \label{sec:Discussion}

\subsection{Observations of Neptune Obliquities}

\begin{figure*}
    \centering
    \includegraphics[width=0.9\linewidth]{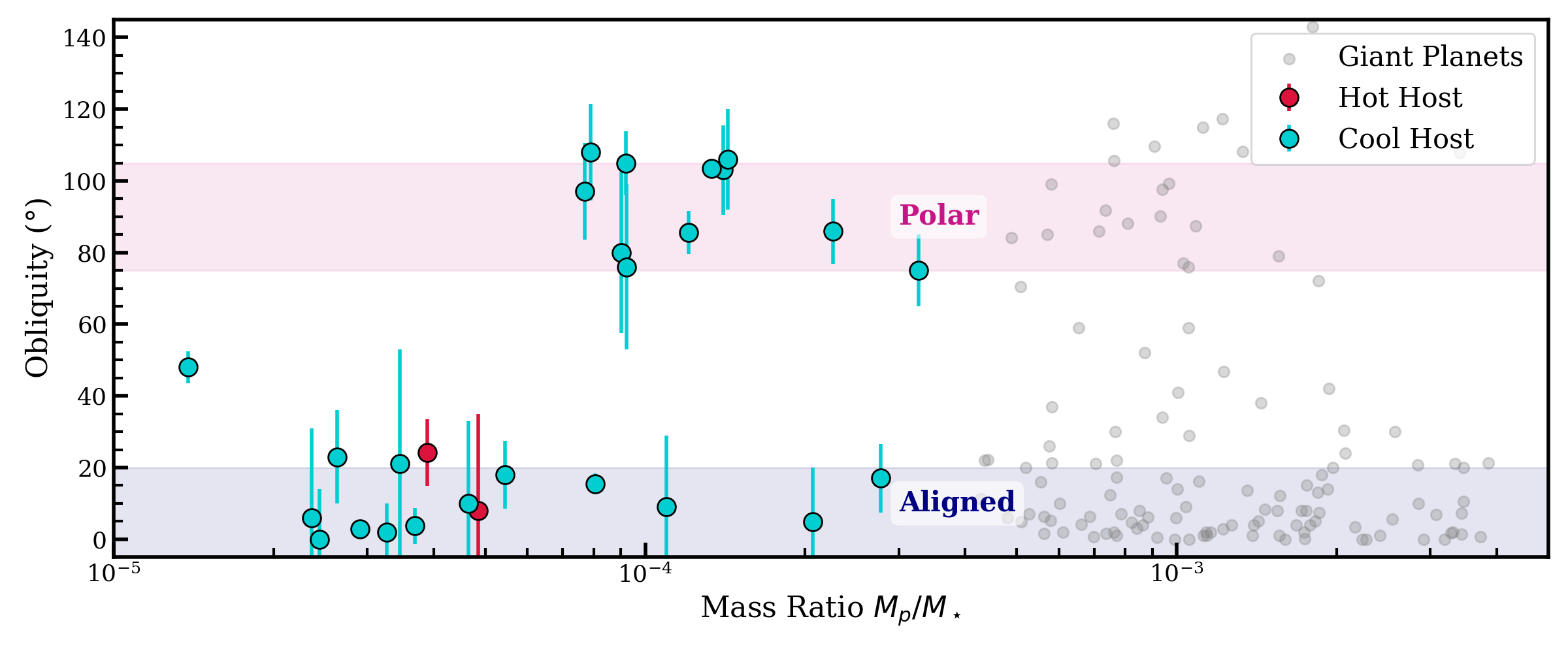}
    \caption{Observations of the stellar obliquity of planets from the TEPCAT database plotted as a function of planet-to-star mass ratio. Here, we focus on low mass hot Neptune planets (colored points) and giant planets (gray points), neglecting planets of masses which bridge the gap and may be difficult to distinguish. While hot ($\gtrsim6200$K) stars are differentiated for Neptune systems in this figure, a qualitative difference from cool hosts is not yet apparent.}
    \label{fig:neptuneobservations}
\end{figure*}

Observations of the RM effect are used to infer the sky-projection of the stellar obliquity angle, $\lambda$, which has proven successful for hundreds of transiting exoplanets. In cases where the rotation period of the star is known (often constrained from periodic spot modulations in the out-of-transit photometric timeseries), the true stellar obliquity $\psi$ may also be constrained \citep{Masuda2020}. We queried the TEPCAT database from \cite{Southworth2011} for measured stellar obliquities of short-period planets ($P<15$ days) and cross-matched targets with the NASA Exoplanet Archive\footnote{\url{https://exoplanetarchive.ipac.caltech.edu/}} to get the planetary parameters. We plot the obliquity measurements (we prefer $\psi$ instead of $\lambda$ whenever available) in Figure \ref{fig:neptuneobservations} as functions of planet-to-star mass ratio for both the $\sim$Neptune mass planets as well as giant planets. Among the polar Neptunes, 11/12 have orbits between 2 and 6 days. This coincides with the `Neptunian Ridge', a relative overabundance of planets of this mass regime and orbital period, noted by \cite{CastroGonzalez2024}.

In our model, we assumed the star aligned with the outer disk such that the inclination state of the Neptune $i_N$ is synonymous with the true obliquity $\psi$. The tendency for Neptunes to cluster around aligned or polar orientations can be easily explained if resonant encounters are often adiabatic and the stellar quadrupole induces greater precession on the planets than the inner disk. If this is characteristic of the typical hot Neptune system, then an aligned-polar dichotomy is a \textit{natural outcome} from our model. Planets with orbits that precess too rapidly due to $J_2$ ($\nu_N^\star>\nu_D$, see Figure \ref{fig:integrations}) will retain aligned architectures, while those that undergo resonance crossing will be excited to a highly inclined (oblique) orbit. Furthermore, the prevalence of polar orbits primarily at short orbital periods can be explained by resonance detuning via ZKL cycles for orbits in which general relativistic precession is not dominant \citep{Petrovich2020}.

\subsection{Prospects for Validation}

There are several observables which can be attributed to the mechanism we have proposed. Most importantly, the dichotomy of polar and aligned orbits should not be sensitive to system ages, at least for those observed post dispersal at $~\gtrsim$10 Myr. Resolving the stellar obliquities of small planets as functions of system age may help to isolate excitation mechanisms with distinct timescales, which has proven successful in a similar problem of the time-dependent disruption of mean-motion resonances \citep{Dai2020,Dai2024}. 

Second, massive planets cannot be preferentially excited to polar orientations due to their substantial back-reactions onto the disk. \cite{Zanazzi2018} discussed this distinction in the context of obliquity excitation in binary systems (although they considered the excitation of the stellar inclination as opposed to that of the planet). The polar end-state is unique to planets which do not strongly influence the cumulative dynamics of the system. Because massive stars have more massive disks (e.g., \cite{Andrews2013}, \cite{Trapman2025}), the maximum mass threshold would also increase around hotter stars. However, there are more complicated processes to consider within such disks, so our predictions are best applied to Sun-like stars.

Finally, the least massive planets (super-Earths and sub-Neptunes) may not respond to the mechanism due to trapping within the disk. Neptune resides in the mass regime for which gap-opening is possible (e.g., \cite{Sanchez2025}). The thermal mass given by solving
\begin{equation}
    \left(\frac{m_\mathrm{thermal}}{3M}\right)^{1/3}\sim \frac{h}{r}
\end{equation}
indicates that at a mass ratio of $\sim$10$^{-4}$ (roughly that of Neptune around a Sun-like star), the planet's Hill sphere swallows the local disk. Even if the inner disk obeys a surface density profile which is amenable to satisfying Equation \ref{eq:crossing}, despite the inclination boost from secular resonance, Type I damping could prevent smaller planets from being excited. If it is indeed the case that Neptunes can reside inside the magnetospheric cavity, a similar argument applies---the gap clearing nature of $\sim$Neptune mass planets may resist the expansion of the cavity, while smaller planets undergo outward migration and remain bound to the disk \citep{Pan2025}. Consistent with this picture, a tentative preference for alignment in the smallest observed planetary systems was noted in \cite{Handley2025} and \cite{Polanski2025}.

We also note that under this scenario, any exterior sub-thermal planets in hot Neptune systems should remain aligned with the disk plane, and therefore be highly mutually inclined relative to the Neptunes. These planets could have thus far eluded transit or RV detection, but our model is not sensitive to their presence. Because super-Earths contribute negligibly to the disk potential (the disk has mass comparable to that of Jupiter), Type I damping would keep them bound to the disk, and they would not qualitatively alter the resonant mechanism. At present, however, such planets have not been identified in the systems of interest.

\subsection{Caveats and Room for Further Work}
\label{sec:caveats}
In this section, we discuss the limitations of our model and analysis, and provide ideas for future work. A few points to consider are: 

\begin{itemize}
    \item Although our analytic model places the planet within the inner cavity of the disk---a configuration that may not be common in practice---we argue that it nonetheless provides a useful illustration of the underlying dynamics for two reasons.
    First, a simple power law surface density profile (which we chose for analytic simplicity) overestimates the surface density in the inner regions of the disk, which is not reflective of the densities in magneto-hydrodynamic simulations. At small radii in the disk, complex turbulent processes dramatically enhance the viscosity, leading to an exponentially diminished surface density \citep{Suzuki2010,Suzuki2016}. Moreover, disks where angular momentum transport is dominated by magneto-centrifugal winds tend to have rather low values of $\Sigma$ at small radii \citep{Pascucci2022}. Rather than model the inner $\sim$0.1~au, we truncate the disk early to capture the dynamics from the dynamically dominant region of the inner disk.
    Second, our Laplace-Lagrange prescription becomes cumbersome for modeling the mutual interaction between the Neptune and the surrounding disk if it is embedded. Evaluating the Laplace coefficients for overlapping orbits leads to unphysical values of $\nu_N^{D,1}$ and $\nu_N^{D,2}$ which could inhibit the resonance mechanism in numerical simulations \citep{Sefilian2025,Lithwick2025}. This issue would persist if we applied an exponential suppression to the surface density to match that of simulations, not only disrupting the true dynamics, but making our analytic expressions more convoluted.


    \item We adopt a low disk viscosity of $\alpha \sim 10^{-4}$, which satisfies the adiabatic criterion (Equation \ref{eq:adiabatic}) for very small mutual inclinations. While higher viscosities could violate the adiabatic criterion, ALMA observations indicate that $\alpha \sim 10^{-4}$ is more consistent among observed disks than earlier estimates of $\alpha \sim 10^{-2}$ \citep{Flaherty2020, Villenave2022, Pizzati2023}.

    \item We assumed a scale for the surface density which is comparable to that of the Minimum-Mass Solar Nebula (MMSN), but our mechanism operates in the photo-evaporative epoch. The surface density of the inner disk could be diminished at Myr timescales, which would present an issue for our assumed angular momentum hierarchy. However, the structure of the inner disk remains uncertain, as sub-au regions are difficult to probe and available diagnostics indicate that inner disks are not universally depleted. In disks where angular momentum removal is dominated by magnetized winds, accretion proceeds through magnetically coupled surface layers and is only weakly tied to the midplane mass reservoir, allowing surface densities near $\sim$1~au to persist over long timescales compared to turbulent ($\alpha$-disk) models \citep{Bai2016,Lesur2021}. Independent support for weak turbulent transport persisting late into disk evolution comes from Solar System constraints, where the non-carbonaceous–carbonaceous (NC-CC) meteorite dichotomy reveals low effective $\alpha$ values and a sustained mass reservoir in this region at several Myr (e.g., \cite{Kleine2020}). Although that structure may not be universal, the hierarchy requirement may naturally produce system-to-system diversity, with only sufficiently massive inner disks evolving toward polar configurations. 

    \item If the gap opens far from the star, the assumption of disk rigidity becomes tenuous. Resonance crossings require the outermost regions of the inner disk to maintain coherence due to bending wave propagation. For a constant aspect ratio, the wave crossing time scales roughly as $\Omega_\mathrm{orb}^{-1}\sim r^{3/2}$, implying that disk warping can become a concern at gaps of several au or more. Minor warping of the disk would not alter the dynamics, so we considered that limit in this work.
\end{itemize}


\section{Conclusion} \label{sec:Conclusion}

Using only the evolution of a protoplanetary disk undergoing photo-evaporation, we have proposed a mechanism which can naturally generate highly inclined orbits for Neptune-mass planets, which manifest as polar spin-orbit angles if observed with the RM effect. Our model is the first proposed mechanism that does so without a giant planet or stellar binary. Using an extended Laplace-Lagrange formalism, we derived an analytic prescription for the disk dispersal phase that reduces to a fundamental resonance model. For a realistic choice of disk parameters, we showed that capture into resonance is easily achieved for disk misalignments of $\sim$1$\degree$. 

Observations of spin-orbit angles for Neptune mass planets are in broad agreement with our model, but the leading prospect for validation is the misalignment of young systems (several Myr) which are not predicted by longer timescale dynamical models. If such primordial tilts are confirmed, they would challenge long-held expectations of coplanarity, and reveal that disk substructures imprint more complexity on planetary architectures than previously assumed.



\begin{acknowledgments}

L.B.H. acknowledges support from the National Science Foundation through the Graduate Research Fellowship Program under Grant No. 2139433. K.B. is thankful for the support of the David and Lucile Packard Foundation, as well as Caltech and 3CPE.

\end{acknowledgments}

\software{
  Mathematica \citep{Mathematica},
  Python \citep{Python},
  NumPy \citep{NumPy},
  Matplotlib \citep{Matplotlib}.
}


\bibliography{main}{}
\bibliographystyle{aasjournalv7}



\end{document}